\documentclass[twocolumn,english,prb,aps]{revtex4-1}
\usepackage[T1]{fontenc}
\usepackage[latin9]{inputenc}
\usepackage{verbatim}
\usepackage{amsmath}
\usepackage{amssymb}
\usepackage{graphicx}
\usepackage{esint}

\makeatletter

\providecommand{\tabularnewline}{\\}

\@ifundefined{textcolor}{}
{%
 \definecolor{BLACK}{gray}{0}
 \definecolor{WHITE}{gray}{1}
 \definecolor{RED}{rgb}{1,0,0}
 \definecolor{GREEN}{rgb}{0,1,0}
 \definecolor{BLUE}{rgb}{0,0,1}
 \definecolor{CYAN}{cmyk}{1,0,0,0}
 \definecolor{MAGENTA}{cmyk}{0,1,0,0}
 \definecolor{YELLOW}{cmyk}{0,0,1,0}
}

\@ifundefined{showcaptionsetup}{}{%
 \PassOptionsToPackage{caption=false}{subfig}}
\usepackage{subfig}
\makeatother

\usepackage{babel}
\begin{document}

\title{Time-reversal invariant topological superconductivity in doped Weyl
semimetals}

\author{Pavan Hosur$^{1}$, Xi Dai$^{2}$, Zhong Fang$^{2}$ and Xiao-Liang
Qi$^{1}$}

\affiliation{$^{1}$Department of Physics, Stanford University\\
 $^{2}$Institute of Physics, Chinese Academy of Sciences}
\begin{abstract}
Time-reversal invariant topological superconductors are a new state
of matter which have a bulk superconducting gap and robust Majorana
fermion surface states. These have not yet been realized in solid
state systems. In this paper, we propose that this state can be realized
in doped Weyl semimetals or Weyl metals. The Fermi surfaces of a Weyl
metal carry Chern numbers, which is a required ingredient for such
a topological superconductor. By applying the fluctuation-exchange
approach to a generic model of time-reversal invariant Dirac
and Weyl semimetals, we investigate what microscopic interactions
can supply the other ingredient, viz., sign changing of the superconducting
gap function between Fermi surfaces with opposite Chern numbers. We
find that if the normal state is inversion symmetric, onsite repulsive
and exchange interactions induce various nodal phases as well as a
small region of topological superconductivity on the phase diagram.
Unlike the He$^{3}$B topological superconductor, the phase here does
not rely on any special momentum dependence of the pairing amplitude.
Breaking inversion symmetry precludes some of the nodal phases and
the topological superconductor becomes much more prominent, especially
at large ferromagnetic interaction. Our approach can be extended to
generic Dirac or Weyl metals. 
\end{abstract}
\maketitle

\section{Introduction}

Nearly a decade after topological insulators took the condensed matter
community by storm, interest in topological band structures is bifurcating
into two main directions. The first is towards topological superconductors
(TSCs)\cite{KitaevClassification,KitaevWireMajorana01,FuBergTopoSC10,QiSCZhangTITSCReview,SchnyderAndreevSurfaceTSC},
which are close cousins of topological insulators from a theoretical
point of view but are a novel phase nonetheless. They share several
properties with the insulators such as a gapped bulk with non-trivial
winding of the wavefunction of the occupied bands intimately tied
to robust, gapless surface states. However, the surface states in
the TSCs are composed of Majorana fermions, in contrast to ordinary
electrons in the insulating counterparts. A well-known example of
a three-dimensional (3D) TSC is the B-phase of He$^{3}$, in which
fermionic He$^{3}$ atoms condense into a superconducting state whose
gap function has a non-trivial texture in momentum space, thus rendering
the superconductor topological\cite{VolovikBook}. However, there
is no known example of a 3D TSC in solid state systems.

The second direction in which interest in topological band structures
is heading is towards gapless systems, ushered forth by the discovery
of a new 3D phase of matter, dubbed \emph{Weyl semimetals} (WSMs).
In this phase, the low-energy electrons behave like Weyl fermions
-- massless, two-component fermions well-known in high-energy physics
and described by the Weyl Hamiltonian $H_{\mbox{Weyl}}\equiv\hbar v\boldsymbol{k}\cdot\boldsymbol{\sigma}$,
where $\sigma_{x,y,z}$ are the three $2\times2$ Pauli matrices and
$\boldsymbol{k}$ is momentum\cite{PyrochloreWeyl,VafekDiracReview,TurnerTopPhases,KrempaWeyl,ChenIridate,HosurWeylReview}.
WSMs can be thought of as a 3D version of graphene; however, unlike
the Dirac nodes in graphene which can be gapped out by breaking point
group symmetries of the honeycomb lattice, Weyl nodes are stable and
can only be annihilated via internode scattering or via superconductivity.
Thus, translational symmetry and charge conservation together render
each Weyl node stable against all symmetry preserving perturbations.
This topological feature of their band structure endows WSMs with
a host of exotic physical properties, ranging from surface states
that form Fermi arcs rather than Fermi surfaces (FSs)\cite{VolovikFlatBands,HosurFermiArcs},
to unusual transport properties hinged on a 3D axial anomaly proportional
to the electromagnetic field $\boldsymbol{E}\cdot\boldsymbol{B}$\cite{AjiABJAnomaly,NielsenABJ,ZyuninBurkovWeylTheta,SonSpivakWeylAnomaly,HosurWeylReview,GrushinWeyl,ParameswaranNonLocalTransport,BasarTriangleAnomaly,LandsteinerAnomaly,QiWeylAnomaly,WeylCDW}.

Simply put, a WSM emerges when a pair of non-degenerate bands intersects
at arbitrary points in momentum space, known as the Weyl nodes. Each
Weyl node can be assigned a handedness -- right or left -- or a \emph{chirality}
quantum\emph{ }number\emph{ $\chi=\pm1$}; the fermion doubling theorem
forces Weyl nodes to always come in pairs with opposite chirality\cite{NielsenFermionDoubling1,NielsenFermionDoubling2}.
Furthermore, even-ness of chirality under time-reversal ($\mathcal{T}$)
ensures that Weyl nodes in a $\mathcal{T}$-invariant WSM occur in
multiples of four. Since the bands must be non-degenerate, inversion
($\mathcal{I}$) symmetry is necessarily broken in such a WSM in order
to respect Kramer's theorem. A general WSM breaks both $\mathcal{T}$
and $\mathcal{I}$ symmetries and Weyl points may be at different
energies, thus turning the system into a Weyl metal phase with a non-vanishing
FS. %

In this work, we show that $\mathcal{T}$-invariant Weyl metals are
natural hosts for realizing a $\mathcal{T}$-invariant TSC, and determine
the microscopic interactions that cause them to do so. That Weyl metals
are convenient starting points for obtaining a TSC can be seen as
follows. In a Weyl metal, the FSs carry non-zero Chern numbers of
the Berry's phase gauge field. The Chern number on the FS encompassing
the $j^{\mbox{th}}$ node, given by 
\begin{eqnarray}
C_{j} & = & \frac{1}{2\pi}\oint_{\mbox{FS}_{j}}\mathrm{d}^{2}k\boldsymbol{\hat{n}}_{j}(\boldsymbol{k})\cdot\boldsymbol{\nabla_{k}}\times{\bf a}_{j}(\boldsymbol{k})\label{Cherndef}
\end{eqnarray}
where ${\bf a}_{j}(\boldsymbol{k})=-i\left\langle \boldsymbol{k},j\right|\boldsymbol{\nabla_{k}}\left|\boldsymbol{k},j\right\rangle $
is the Berry connection for the state $\left|\boldsymbol{k},j\right\rangle $
on the $j^{\mbox{th}}$ FS and $\boldsymbol{\hat{n}}_{j}(\boldsymbol{k})=\boldsymbol{v}_{j}^{F}(\boldsymbol{k})/\left|\boldsymbol{v}_{j}^{F}(\boldsymbol{k})\right|$
is the FS normal, equals the chirality of the node: $C_{j}=\chi_{j}=\pm1$
irrespective of the sign of the doping. In Ref \onlinecite{QiFSInvariantTSC},
a simple formula was discovered relating the FS Chern number to topological
superconductivity: given a $\mathcal{T}$-invariant metal with a set
of FSs with Chern numbers $\{C_{j}\}$, it was shown that the topological
invariant $\nu$ for a $\mathcal{T}$-invariant TSC is 
\begin{equation}
\nu=\frac{1}{2}\sum_{j\in\mbox{FS}}C_{j}\mbox{sgn}\left(\Delta_{j}\right)\label{topoinvariant}
\end{equation}
where $\Delta_{j}$ is the pairing gap function on the $j^{\mbox{th}}$
FS and is assumed to be much smaller than the Fermi energy in magnitude.
$\mathcal{T}$-symmetry ensures that $\Delta_{j}$ is real, and sgn$(\Delta_{j})$
is well-defined because of the requirement of a fully gapped state.
A TSC is implied by $\nu\neq0$; thus, doped WSMs are natural parent
compounds for realizing such a phase since the nontrivial Chern number
is already provided by the band structure. The superconducting order
parameter need not have a specific momentum-dependence like in He$^{3}$B.
Instead, it can be momentum-independent on each FS but must alternate
in sign between different FSs in any pattern that makes the weighted
sum (\ref{topoinvariant}) non-vanishing. The question that now begs
to be answered is: what microscopic interactions will induce appropriate
sign changes in the pairing gap so that the upshot is a TSC?

We answer this question by computing the pairing instabilities in
a generic model for a $\mathcal{T}$-invariant WSM using the
fluctuation-exchange approach\cite{ScalapinoDWave,RaghuWeakRepulsiveSC}.
The pairing instabilities are generically the eigenstates of the effective
interaction vertex at the Fermi level, $\chi_{\mathcal{T}}^{ij}(\boldsymbol{k},\boldsymbol{k}')$,
which is defined as the amplitude for two-particle scattering from
a pair of Kramer's conjugates $|\boldsymbol{k},i\rangle\otimes|\hat{\mathcal{T}}(\boldsymbol{k},i)\rangle$
to another pair $|\boldsymbol{k}',j\rangle\otimes|\hat{\mathcal{T}}(\boldsymbol{k}',j)\rangle$.
To first order in the interactions, the vertex is simply the projection
of the bare interaction onto the FS states. Within the fluctuation-exchange
prescription, it is obtained to higher orders perturbatively by integrating
out states away from the Fermi level. Once $\chi_{\mathcal{T}}^{ij}(\boldsymbol{k},\boldsymbol{k}')$
is determined, the dominant pairing instability is given by the most
negative eigenvalue $\lambda_{T}$ of the linearized gap equation
($\hbar=1$ henceforth) 
\begin{equation}
\sum_{j}\intop_{\boldsymbol{k}'} \delta(v_{j}k'-|\mu_{j}|)\chi_{\mathcal{T}}^{ij}(\boldsymbol{k},\boldsymbol{k}')\Delta_{j}(\boldsymbol{k}')=\lambda_{T}\Delta_{i}(\boldsymbol{k})\label{eq:eigenvalue-equation}
\end{equation}
where $\mu_{j}$ is the chemical potential relative to the $j^{\mbox{th}}$
Weyl point and $\intop_{\boldsymbol{k}'}\equiv \int\frac{\mathrm{d}^3k'}{(2\pi)^3}$, and the gap function $\Delta_{j}(\boldsymbol{k})$ is
given by the corresponding eigenvector. If all the eigenvalues are
non-negative, there is no pairing instability. When there \emph{is}
an instability, the critical temperature for the phase transition
is of the form $T_{c}\sim\Lambda e^{-1/\rho\left|\lambda_{T}\right|}$
where $\Lambda$ is an energy cutoff and $\rho$ is the density of
states at the Fermi level. Eq. (\ref{eq:eigenvalue-equation}) is
essentially the statement that the broken symmetry phase that ultimately
forms is given by the global minimum of the free energy and corresponds
to its most negative eigenvalue.

Physically, a pairing instability exists in $\mathcal{T}$-symmetric
systems because \emph{every} state has a degenerate Kramer's partner
with which it can form a Cooper pair and fall into a coherent condensate
in the presence of attractive interactions. In the language of renormalization,
this means that only attractive interactions between pairs of Kramer's
conjugates states are (marginally) relevant, while all other interactions
are either irrelevant or have no flow due to severe phase space constraints.
We shall label these Cooper pairs `type-T'. In $\mathcal{I}$-symmetric
systems, a second type -- `type-I' -- of Cooper pairs is also possible
where the electrons in the pair are related by $\mathcal{I}$. The
corresponding pairing states are given by the eigenstates of an
effective interaction matrix $\chi_{\mathcal{I}}^{ij}(\boldsymbol{k},\boldsymbol{k}')$
analogous to the interaction for type-T states: 
\begin{equation}
\sum_{j}\intop_{\boldsymbol{k}'}{}\delta(v_{j}k'-|\mu_{j}|)\chi_{\mathcal{I}}^{ij}(\boldsymbol{k},\boldsymbol{k}')\Delta_{j}(\boldsymbol{k}')=\lambda_{I}\Delta_{i}(\boldsymbol{k})\label{eq:eigenvalue-equation-1}
\end{equation}
The leading pairing instability in $\mathcal{I}$-symmetric systems
is then given by the combined lowest eigenvalue of $\chi_{\mathcal{I}}$
and $\chi_{\mathcal{T}}$. More generally, superpositions of type-T
and type-I Cooper pairs can exist in systems with both $\mathcal{T}$-
and $\mathcal{I}$-symmetries. However, we will see later that such
mixing is forbidden in our system by symmetry.

Applying the above prescription, for concreteness, on the Hamiltonian
for a known Dirac semimetal Na$_{3}$Bi\cite{WangA3Bi} and restricting
to onsite interactions for simplicity, we find that the TSC forms
over a large part of the phase diagram, especially when $\mathcal{I}$
symmetry is broken. In particular, this occurs for purely Ising ferromagnetic
interactions ($J_{z}<0$) and survives moderate values of Hubbard
($U>0$) and inter-orbital ($V>0$) repulsion. If $\mathcal{I}$-symmetry
is restored, most of the TSC is overwhelmed by a nodal phase and only
a narrow region near the $V=-J_{z}$ line survives. Along the way,
we unearth various other nodal phases with point or line nodes. These
results are summarized in Figs. \ref{fig:Phase-diagram-second-order-I}
and \ref{fig:Phase-diagram-second-order-T} and in Table \ref{tab:Phases}.
We emphasize that although we start with a particular model for Dirac
semimetals, our results can straightforwardly be extended to other
models by simply reinterpreting the orbital content of the Dirac matrices.

Our work complements two recent works on superconducting instabilities
of doped WSMs\cite{ChoWeylSC,WeiWeylSC}. Both these works consider
$\mathcal{I}$-symmetric WSMs, which necessarily break $\mathcal{T}$-symmetry,
and study superconducting phases within mean field theory. In contrast,
we focus on $\mathcal{T}$-symmetric WSMs and compute the pairing
instabilities using an approach that is more unbiased than mean field
theory.

\section{Necessary conditions for the TSC\label{sec:microscopic-origins}}

Before presenting the detailed calculation on our model system, we
develop some general intuition on the microscopic origins of the TSC.
Consider a ``minimal'' $\mathcal{T}$-symmetric Weyl metal with
four isotropic Weyl nodes, 
\begin{equation}
\mathcal{H}_{j}(\boldsymbol{k})=(-1)^{j}\hbar v_{|j|}\boldsymbol{k}\cdot\boldsymbol{\Gamma}_{j}-\mu_{|j|},\, j\in\{1,-1,2,-2\}
\end{equation}
where $\boldsymbol{\Gamma}_{j}$ are $2\times2$ Pauli matrices in
the basis of the local degrees of freedom at the $j^{\mbox{th}}$
Weyl node, and $\boldsymbol{k}$ is the momentum measured with respect
to the Weyl point wavevector. The corresponding energy eigenvalues
are $\mathcal{E}_{j}^{\pm}(\boldsymbol{k})=\mbox{sgn}\left(\mu_{|j|}\right)\left(\hbar v_{|j|}k\pm\left|\mu_{|j|}\right|\right)$.
$\mathcal{H}_{\pm j}$ are related by $\mathcal{T}$ and have the
same FS Chern number $C_{j}=(-1)^{j}$. Isotropy of each Weyl node
can be assumed without loss of generality as any anisotropy can be
removed by locally rescaling momentum relative to the Weyl node.

\begin{figure}

\begin{centering}
\includegraphics[width=0.7\columnwidth]{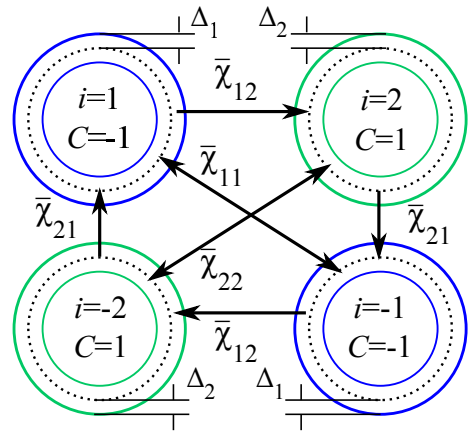} 
\end{centering}

\caption{Schematic illustration of the pairing interaction and the gap functions
described in Sec \ref{sec:microscopic-origins}. The dotted circles
denote the normal state FSs with index $i$ as discussed in the text
and Chern numbers $C=\pm1$, while the solid lines represent the gaps
$\Delta_{1,2}$ in the superconducting state. $\bar{\chi}_{ij}$ is
the average Cooper scattering amplitude from Kramer's conjugate states
on the $(i,-i)$ FSs to the $(j,-j)$ FSs.\label{fig:four-FSs}}

\end{figure}

For isotropic Weyl nodes, the superconducting gap function projected
onto the FSs in any $\mathcal{T}$-symmetric gapped phase must have
the form $\Delta_{j}(\boldsymbol{k})=\Delta_{|j|}$, i.e., it must
be independent of $\boldsymbol{k}$ and the same for Kramer's conjugate
FSs. $\mathcal{T}$-symmetry further requires $\Delta_{|j|}\in\mathbb{R}$.
For weak pairing, the Bogoliubov-de Gennes Hamiltonian that describes
pairing of Kramer's conjugate Weyl nodes can be written as 
\begin{equation}
H_{j}^{BdG}=\Psi_{j}^{\dagger}(\boldsymbol{k})\left(\begin{array}{cc}
\mathcal{E}_{j}^{-}(\boldsymbol{k}) & \Delta_{|j|}\\
\Delta_{|j|} & -\mathcal{E}_{j}^{-}(\boldsymbol{k})
\end{array}\right)\Psi_{j}(\boldsymbol{k})
\end{equation}
where $\Psi_{j}^{\dagger}(\boldsymbol{k})=\left(\psi_{j}^{\dagger}(\boldsymbol{k}),\psi_{-j}(-\boldsymbol{k})\right)$
is the usual Nambu spinor corresponding to $\psi_{j}^{\dagger}(\boldsymbol{k})$,
the creation operator for a fermion at the Fermi level at momentum
$\boldsymbol{k}$ relative to the $j^{\mbox{th}}$ Weyl node. The
form $\Delta_{j}(\boldsymbol{k})=\Delta_{|j|}\in\mathbb{R}$ block
diagonalizes $\chi_{ij}$ and simplifies (\ref{eq:eigenvalue-equation})
to 
\begin{equation}
\left(\begin{array}{cc}
\bar{\chi}_{11} & \bar{\chi}_{12}\\
\bar{\chi}_{12} & \bar{\chi}_{22}
\end{array}\right)\left(\begin{array}{c}
\Delta_{1}\\
\Delta_{2}
\end{array}\right)=\lambda_{T}\left(\begin{array}{c}
\Delta_{1}\\
\Delta_{2}
\end{array}\right)\label{eq:avg-eigenvalue-equation}
\end{equation}
where $\bar{\chi}_{ij}=\bar{\chi}_{ji}=\bar{\chi}_{ij}^{*}$ is $\chi_{\mathcal{T}}^{ij}(\boldsymbol{k},\boldsymbol{k}')$
averaged over $\boldsymbol{k}$ and $\boldsymbol{k}'$ on FSs $i$
and $j$, respectively: 
\begin{equation}
\bar{\chi}_{ij}=\intop_{\boldsymbol{k},\boldsymbol{k}'}\delta\left(v_{|i|}k-\left|\mu_{i}\right|\right)\delta\left(v_{|j|}k^{\prime}-\left|\mu_{j}\right|\right)\chi_{\mathcal{T}}^{ij}(\boldsymbol{k},\boldsymbol{k}')\label{eq:chi-avg}
\end{equation}
as depicted in Fig \ref{fig:four-FSs}. We have used fermion antisymmetry
in conjunction with $\hat{\mathcal{T}}^{2}=-1$ in deriving (\ref{eq:avg-eigenvalue-equation}).
The topological invariant defined in (\ref{topoinvariant}) then reduces
to 
\begin{eqnarray}
\nu & = & \frac{1}{2}\left(2{\rm sgn}\left(\Delta_{1}\right)-2{\rm sgn}\left(\Delta_{2}\right)\right)\nonumber \\
 & = & {\rm sgn}\left(\Delta_{1}\right)-{\rm sgn}\left(\Delta_{2}\right)
\end{eqnarray}
Therefore the necessary condition for the TSC is that $\Delta_{1}$
and $\Delta_{2}$ have opposite signs. For the TSC to be favored over the trivial state, in which $\Delta_1$ and $\Delta_2$ have the same sign, the criteria $\bar{\chi}_{ij}$ must fulfil are
\begin{equation}
\bar{\chi}_{12}>0\mbox{ and }\begin{cases}
\bar{\chi}_{11}+\bar{\chi}_{22}<0 & \mbox{ or}\\
\bar{\chi}_{11}\bar{\chi}_{22}<\left(\bar{\chi}_{12}\right)^{2}
\end{cases}\label{eq:necessary-conditions}
\end{equation}

These requirements are quite non-trivial. Purely attractive effective
interactions violate the first condition and instead give a trivial
superconductor. On the other hand, purely repulsive ones satisfy (\ref{eq:necessary-conditions})
only if inter-FS scattering is stronger than intra-FS scattering,
which is unnatural. One way to satisfy these conditions is by including
Coulomb repulsion as well as attractive interactions, such as those
mediated by phonons, and fine-tuning their relative strengths. This
way, the net interaction can be made to change sign over large momenta,
comparable to the separation of the Weyl nodes, as required by (\ref{eq:necessary-conditions}).
It would, however, be nicer if the necessary momentum dependence emerged
naturally without fine-tuning. Note that (\ref{eq:necessary-conditions})
are not sufficient conditions for a TSC; they ensure that the TSC
wins over the trivial superconductor but do not rule out nodal phases.

Eq. (\ref{eq:necessary-conditions}) are conditions on the effective
interactions at the Fermi level. Next, we ask, can an onsite interaction
$\mathcal{U}^{\alpha\beta\gamma\delta}\alpha^{\dagger}\beta^{\dagger}\gamma\delta$,
assumed to be $\mathcal{T}$-symmetric, give effective interactions
that satisfy (\ref{eq:necessary-conditions})? Naively, this seems
impossible because (\ref{eq:necessary-conditions}) requires interactions
to depend on the FS indices and hence, on momentum, but onsite interactions
are momentum-independent%
\footnote{The FS index is also a momentum index because it labels Weyl nodes
which are generally separated in momentum space.%
}. However, the effective interaction can acquire momentum dependence
in two ways.

Firstly, the FS wavefunctions and hence the operators that project
the interactions onto the FSs depend on momentum. That is, projecting
$\mathcal{U}$ onto Kramer's conjugate states on the FSs gives $\chi$:
\begin{equation}
\chi_{\mathcal{T}}^{ij}(\boldsymbol{k},\boldsymbol{k}')=\mathcal{U}^{\alpha\beta\gamma\delta}\langle\boldsymbol{k}',j|\alpha\rangle\langle\hat{\mathcal{T}}(\boldsymbol{k}',j)|\beta\rangle\langle\gamma|\hat{\mathcal{T}}(\boldsymbol{k},i)\rangle\langle\delta|\boldsymbol{k},i\rangle
\end{equation}
which is momentum dependent. However, it turns out that such momentum
dependence cannot give a gapped TSC. To see this, we
rewrite the $\boldsymbol{k}'$-dependent part of the right hand side above as
\begin{eqnarray}
\langle\boldsymbol{k}',j|\alpha\rangle\langle\hat{\mathcal{T}}(\boldsymbol{k}',j)|\beta\rangle & = & -\langle\hat{\mathcal{T}}\beta|\boldsymbol{k}',j\rangle\langle\boldsymbol{k}',j|\alpha\rangle\nonumber \\
 & = & -\left\langle \hat{\mathcal{T}}\beta\left|P_{j}(\boldsymbol{k}')\right|\alpha\right\rangle ,
\end{eqnarray}
with $P_{j}(\boldsymbol{k}')=|\boldsymbol{k}',j\rangle\langle\boldsymbol{k}',j|=\frac{1}{2}\left(1+\frac{(-1)^{j}\boldsymbol{k}'\cdot{\bf \Gamma_{j}}}{k'}\right)$
the projection operator onto state $|\boldsymbol{k}',j\rangle$. Clearly,
the $j$-dependent part of $P_{j}(\boldsymbol{k}')$ vanishes after
integration over $\boldsymbol{k}^{\prime}$. Therefore, $\bar{\chi}_{\mathcal{T}}^{ij}$
obtained by a FS average of $\chi_{\mathcal{T}}^{ij}(\boldsymbol{k},\boldsymbol{k}')$
is independent of the FS indices and cannot support a TSC. In summary,
assuming only $\mathcal{T}$-symmetry and linear dispersion, bare
on-site interactions cannot satisfy conditions (\ref{eq:necessary-conditions})
for a TSC even though the FS effective interaction is momentum dependent.

Secondly, $\chi_{\mathcal{T}}^{ij}(\boldsymbol{k},\boldsymbol{k}')$
receives contributions from virtual processes at higher orders in
the bare interactions which are, in general, momentum dependent. A
consequence of these induced momentum dependences is that there can
be channels in which the gap function $\Delta$ is also momentum dependent,
in such a way that it sees an effective attractive interaction. In
other words, there can be eigenstates $\Delta(\boldsymbol{k})$ of
(\ref{eq:eigenvalue-equation}) with negative eigenvalue $\lambda$.
In the past, such a procedure has been used to predict non-zero angular
momentum pairing states induced by Hubbard repulsion\cite{ScalapinoDWave,RaghuWeakRepulsiveSC}.
As we will show explicitly in a model system, it is indeed possible
to induce a TSC with effective interactions incorporating such second
order processes.

As a final note before moving onto the actual calculation for a prototype
model, we point out that type-I Cooper pairing in $\mathcal{I}$-symmetric
systems cannot induce topological superconductivity. Since $\mathcal{I}$
relates FSs with opposite Chern numbers, the pair amplitude on each
FS necessarily has a phase that winds around some axis passing through
the corresponding Weyl node, which in turn requires it to have nodes
on the FS. Thus, there are no type-I pairing states that are fully
gapped and hence, there is no topological superconductivity.

\section{A prototype model \label{sec:ABi-system}}

Having proven that higher order processes are necessary for obtaining
a TSC in a general $\mathcal{T}$-symmetric WSM, we now apply our
analysis to a prototype model which describes a $\mathcal{T}$-symmetric
WSM with minimal number of Weyl points for concreteness. We start
from the Dirac semimetal A$_{3}$Bi proposed in Ref. \onlinecite{WangA3Bi},
with A$=$Na, K, Rb. ${\rm Na_{3}Bi}$ has been recently realized
experimentally\cite{xu2013,liu2014}. Although A$_{3}$Bi is a Dirac
semimetal\cite{young2012} with two Dirac nodes, it can be thought
of as a parent compound for a $\mathcal{T}$-invariant WSM with four
Weyl nodes, because the latter can be obtained from the former via a suitable $\mathcal{I}$-breaking
structural deformation. Below, we first use Na$_{3}$Bi as an example
to explain the prototype model with $\mathcal{I}$-symmetry and then
discuss the effects of $\mathcal{I}$-symmetry breaking. Effective
interaction terms will be investigated post the description of the
band structure. All discussion about the effective model applies to
other A$_{3}$Bi materials.

\subsection{The prototype model of Dirac semimetal}

The crystal structure of Na$_{3}$Bi consists of two inequivalent
sets of Na atoms, Na(1) and Na(2) and one set of Bi atoms. Na(1) atoms
form a honeycomb lattice with Bi, while Na(2) are interspersed between
the honeycomb layers. The low energy theory near the Dirac point involves
four orbitals, including an even parity superposition of the 3$s$
orbitals of Na(1) and Na(2), henceforth referred to as $S$, and an
odd-parity superposition of the 6 $p_{x}\pm ip_{y}$ orbitals of Bi
atoms in adjacent honeycomb layers, which we shall call $P$. With
spin-orbit coupling, the low energy bands near Fermi level are $\left|s\uparrow\right\rangle ,\left|p_{x}+ip_{y},\uparrow\right\rangle ,\left|s\downarrow\right\rangle ,\left|p_{x}-ip_{y},\downarrow\right\rangle $
which have angular momentum quantum number $J_{z}=\frac{1}{2},\frac{3}{2},-\frac{1}{2},-\frac{3}{2}$
respectively. The low energy effective Hamiltonian of these four bands
can be written as $H_{0}(\boldsymbol{k})=c_{\boldsymbol{k}}^{\dagger}h_{0}(\boldsymbol{k})c_{\boldsymbol{k}}$
with 
\begin{eqnarray}
h_{0}(\boldsymbol{k}) & = & A(\tau_{x}\sigma_{z}k_{x}-\tau_{y}k_{y})-(M_{0}-M_{1}k_{z}^{2})\tau_{z}\nonumber \\
 &  & +\left(\epsilon(\boldsymbol{k})-\mu\right)\mathbb{I}\nonumber \\
 & = & \left(\begin{array}{cccc}
M(\boldsymbol{k}) & Ak_{+} & 0 & 0\\
Ak_{-} & -M(\boldsymbol{k}) & 0 & 0\\
0 & 0 & M(\boldsymbol{k}) & -Ak_{-}\\
0 & 0 & -Ak_{+} & -M(\boldsymbol{k})
\end{array}\right)\nonumber \\
 &  & +\left(\epsilon(\boldsymbol{k})-\mu\right)\mathbb{I}\label{HDirac}
\end{eqnarray}
Here $c_{\boldsymbol{k}}$ is a four-component fermion annihilation
operator, $\tau_{x,y,z}$ and $\sigma_{x,y,z}$ are Pauli matrices
in the orbital and spin-space, respectively, $A$, $M_{0}$ and $M_{1}$
are positive constants, $\epsilon_{0}(\boldsymbol{k})$ is a $\boldsymbol{k}$-dependent
energy shift and $\mu$ is the chemical potential\cite{WangA3Bi}.
We have denoted $M(\boldsymbol{k})=M_{0}-M_{1}k_{z}^{2}$ in the second line
above. Note the distinction between $c_{\boldsymbol{k}}$ and $\psi_j({\boldsymbol{k})}$
introduced in Sec \ref{sec:microscopic-origins}; the former is a
four-component spinor in the spin and orbital basis whereas the latter
is a one-component field representing a state at the Fermi level.
$h_{0}(\boldsymbol{k})$ has Dirac nodes at $\pm K\hat{\mathbf{z}}=\pm\sqrt{M_{0}/M_{1}}\hat{\mathbf{z}}$.
The symmetries of the system are time reversal $\hat{\mathcal{T}}\equiv i\sigma_{y}\mathbb{K}$,
inversion $\hat{\mathcal{I}}\equiv\tau_{z}$, spin rotation $\hat{\mathcal{S}}_{z}\equiv\sigma_{z}/2$
as well as threefold rotation symmetry about the $c$-axis. Here $\mathbb{K}$
is the complex conjugation operator. In the long wavelength limit,
the threefold rotation symmetry is enlarged to a continuous rotation
symmetry $\mathcal{R}_{z}\equiv-i\partial_{\phi}-\tau_{z}\sigma_{z}/2$
to quadratic order in the momentum, with $-i\partial_{\phi}$ ($-\tau_{z}\sigma_{z}/2$)
the orbital (spin) angular momentum.

\begin{figure}
\begin{centering}
\includegraphics[width=1\columnwidth]{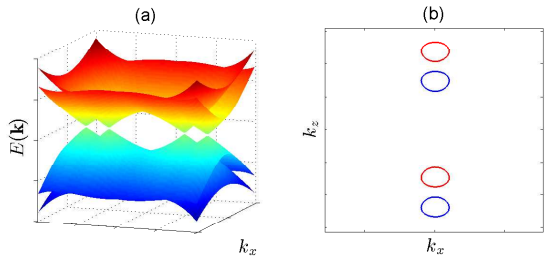} 
\par\end{centering}

\caption{The energy dispersion (a) and the corresponding FS contour (b) in
the $k_{x}-k_{z}$ plane, for the $\mathcal{I}$-breaking term given
by Eq. (\ref{Ibreaking1}). The red and blue contours in (b) correspond
to spin up and down FSs, respectively.\label{fig:weylFS}}
\end{figure}

\subsection{Effects of inversion symmetry breaking}

Adding $\mathcal{I}$ breaking terms lifts the degeneracy between
the two pairs of Kramers degenerate bands at each Dirac point. For
example, a simple term 
\begin{equation}
h_{1}(\boldsymbol{k})=2M_{1}\delta Kk_{z}\tau_{z}\sigma_{z}\label{Ibreaking1}
\end{equation}
splits the two Dirac nodes into four Weyl nodes located at $\left(\pm K\pm\delta K\right)\hat{\mathbf{z}}$
to first order in $\delta K/K$, %
{} while preserving the symmetries $\hat{\mathcal{R}}_{z}$, $\hat{\mathcal{S}}_{z}$
and $\hat{\mathcal{T}}$, as shown in Fig \ref{fig:weylFS}.

$\mathcal{I}$-symmetry breaking has crucial effects on superconductivity.
In the $\mathcal{I}$-symmetric Dirac semimetal, there are two degenerate
states at each wavevector on the FS. Thus, a
state $|\boldsymbol{k},i\rangle$ can
generically form a Cooper pair with a superposition of its time-reversal
partner $|\mathcal{T}\left(\boldsymbol{k},i\right)\rangle$ and inversion
partner $|\mathcal{I}\left(\boldsymbol{k},i\right)\rangle$, i.e., it can form 
a superposition of a type-T and a type-I Cooper pairs. In the current
model, type-T Cooper pairs have total $\mathcal{S}_{z}=0$ while type-I
Cooper pairs have total $\mathcal{S}_{z}=\pm1$. Due to $\mathcal{S}_{z}$-conservation,
there is thus no mixing between the two types of Cooper pairs. On
the other hand, when the two degenerate bands are split by an $\mathcal{I}$-breaking
term, there is only one state $|\boldsymbol{k},i\rangle$ at a generic
point at the Fermi level, and the state at opposite momentum is $|\mathcal{T}(\boldsymbol{k},i)\rangle$.
The inversion partner $|\mathcal{I}(\boldsymbol{k},i)\rangle$ now
has a different energy and moves away from the Fermi level. As long
as the splitting energy scale is much larger than the superconducting
gap, Cooper pairing can only occur between the Kramers partners $|\boldsymbol{k},i\rangle$
and $|\mathcal{T}(\boldsymbol{k},i)\rangle$. In other words, $\mathcal{I}$-symmetry
breaking suppresses some of the pairing channels and thus improves
the stability of the $\mathcal{T}$-invariant superconducting phases,
including the TSC that we seek. Indeed, in our fluctuation-exchange
calculation that will be discussed later, we find that the TSC phase
is much more stable in the absence of $\mathcal{I}$-symmetry.

\subsection{Effective interaction terms}

Next, we discuss interaction terms that are consistent with the symmetries
of the system. For A$_{3}$Bi, the electron orbitals near the Fermi
level are itinerant, so that we can start from the band structure
model discussed above, and treat a general short-ranged interaction
as a perturbation. To the leading order, it leads to a quartic term
in the four-band model: \begin{widetext} 
\begin{eqnarray}
H_{{\rm int}} & = & \sum_{\boldsymbol{k},\boldsymbol{k}',\boldsymbol{q},\alpha\beta\gamma\delta}c_{\boldsymbol{k}+\boldsymbol{q},\alpha}^{\dagger}c_{\boldsymbol{k}'-\boldsymbol{q},\beta}^{\dagger}c_{\boldsymbol{k}',\gamma}c_{\boldsymbol{k},\delta}g_{\alpha\beta\gamma\delta}\left(\boldsymbol{k},\boldsymbol{k}',\boldsymbol{q}\right)\label{Hint}\\
\text{with~}g_{\alpha\beta\gamma\delta}\left(\boldsymbol{k},\boldsymbol{k}',\boldsymbol{q}\right) & = & \int\mathrm{d}^{3}rd^{3}r'u_{\boldsymbol{k}+\boldsymbol{q}\alpha}^{*}(\boldsymbol{r})u_{\boldsymbol{k}'-\boldsymbol{q},\beta}^{*}(\boldsymbol{r})u_{\boldsymbol{k}'\gamma}(\boldsymbol{r})u_{\boldsymbol{k}\delta}(\boldsymbol{r})V^{\alpha\beta\gamma\delta}\left(\boldsymbol{r}-\boldsymbol{r}'\right)e^{i\boldsymbol{q}\cdot(\boldsymbol{r}-\boldsymbol{r}')}\nonumber 
\end{eqnarray}
\end{widetext} Here $u_{\boldsymbol{k}\alpha}(\boldsymbol{r})$ is
the periodic part of the Bloch function for the orbital $\alpha$,
with $\alpha$ running over the four orbitals in the effective model,
and $V^{\alpha\beta\gamma\delta}(\boldsymbol{r}-\boldsymbol{r}')$
is a general short-ranged interaction. Note that there is no sum over
the Greek indices in the second line.

We now make the following two simplifying approximations. Firstly,
in all materials discussed here, the Weyl points are close to the
$\Gamma$ point in the Brillouin zone, so that we can approximate
$u_{\boldsymbol{k}\alpha}(\boldsymbol{r})$ by its $\Gamma$ point
value $u_{\boldsymbol{0}\alpha}(\boldsymbol{r})$. Secondly, the Fourier
transform of short-ranged interaction potential is
smooth in $\boldsymbol{q}$, so that we can expand the interaction
vertex in powers of $\boldsymbol{q}$: $g_{\alpha\beta\gamma\delta}\left(\boldsymbol{k},\boldsymbol{k}',\boldsymbol{q}\right)\simeq g_{\alpha\beta\gamma\delta}\left(\boldsymbol{0},\boldsymbol{0},\boldsymbol{q}\right)\simeq g_{\alpha\beta\gamma\delta}\left(\boldsymbol{0},\boldsymbol{0},\boldsymbol{0}\right)+\left.\boldsymbol{q}\cdot\boldsymbol{\nabla_{q}}g_{\alpha\beta\gamma\delta}\left(\boldsymbol{0},\boldsymbol{0},\boldsymbol{q}\right)\right|_{\boldsymbol{q}=\boldsymbol{0}}+...$.
To leading order in $\boldsymbol{q}$, we approximate the interaction vertex by the
momentum independent form $g_{\alpha\beta\gamma\delta}\left(\boldsymbol{0},\boldsymbol{0},\boldsymbol{0}\right)\equiv g_{0\alpha\beta\gamma\delta}$,
which in real space corresponds to approximating the short-range interaction
by an on-site interaction. In short, the location of the Weyl nodes
close to the $\Gamma$-point allows us to strip the interaction of
any weak intrinsic momentum dependence. This is justified insofar
as the TSC is sought, since a weak momentum dependence cannot change
the sign of the interaction over momenta comparable to the splitting
of the Weyl node, which was shown to be necessary condition for the
TSC in Sec \ref{sec:microscopic-origins}.

Either with the approximations above, or phenomenologically, we can
now write down the generic form of the on-site interaction vertex
$g_{0\alpha\beta\gamma\delta}$ that is consistent with the symmetries
of the system, including $\mathcal{T}$-symmetry, $\mathcal{R}_{z}$-symmetry
and $\mathcal{S}_{z}$-conservation. The allowed terms are:
\begin{enumerate}
\item Density-density interactions:

\begin{enumerate}
\item Hubbard repulsion: $U_{S}n_{S\uparrow}n_{S\downarrow}+U_{P}n_{P\uparrow}n_{P\downarrow}$ 
\item Inter-orbital repulsion: $V(n_{S\uparrow}+n_{S\downarrow})(n_{P\uparrow}+n_{P\downarrow})$ 
\end{enumerate}
\item Ising exchange: $J_{z}S_{i}^{\dagger}\sigma_{z}^{ij}S_{j}P_{k}^{\dagger}\sigma_{z}^{kl}P_{l}$ 
\item Pair-hopping: $W\left(S_{\uparrow}^{\dagger}S_{\downarrow}^{\dagger}P_{\downarrow}P_{\uparrow}+P_{\uparrow}^{\dagger}P_{\downarrow}^{\dagger}S_{\downarrow}S_{\uparrow}\right)$ 
\end{enumerate}
Here $S_{i}=(S_{\uparrow},S_{\downarrow})$ and $P_{i}=(P_{\uparrow},P_{\downarrow})$
are the annihilation operators of the four bands, and $n_{S}=S_{i}^{\dagger}S_{i},~n_{P}=P_{i}^{\dagger}P_{i}$
are the net fermion numbers in the $S$ and $P$ orbitals, respectively.
All these interactions preserve $\mathcal{I}$ symmetry. Moreover,
breaking $\mathcal{I}$ symmetry while preserving the other symmetries
does not allow any more terms. If we assume only the three-fold rotation
symmetry instead of the continuous rotation symmetry $\mathcal{R}_{z}$,
more interaction terms will be allowed. However, the effect of these
terms to the low energy physics is suppressed since the $\mathcal{R}_{z}$-symmetry
breaking terms are of order $k^{3}$. Thus, we will only consider
the four kinds of interactions listed above.

\section{The fluctuation-exchange calculation\label{sec:calculation}}

We now calculate the effective interaction vertices $\chi_{\mathcal{T}}^{ij}(\boldsymbol{k},\boldsymbol{k}')$
and $\chi_{\mathcal{I}}^{ij}(\boldsymbol{k},\boldsymbol{k}')$ for
the prototype model and then use (\ref{eq:eigenvalue-equation}) and
(\ref{eq:eigenvalue-equation-1}) to find the leading pairing instability.
Since we are interested in the physics of the Weyl nodes, we assume
$|\mu|,\delta K\ll K$. In this limit, it is reasonable to linearize
the Hamiltonian $H_{0}$ and work, instead, with 
\begin{eqnarray}
H_{0}(\boldsymbol{k}) & = & c_{\boldsymbol{k}}^{\dagger}\left[\tau_{x}\sigma_{z}k_{x}-\tau_{y}k_{y}+\nu_{z}\tau_{z}k_{z}-\mu\right]c_{\boldsymbol{k}}\label{eq:linearized H}\\
 & \equiv & \sum_{j}c_{\boldsymbol{k},j}^{\dagger}\left(\boldsymbol{\Gamma}_{j}\cdot\boldsymbol{k}-\mu\right)c_{\boldsymbol{k},j}\nonumber 
\end{eqnarray}
where $\Gamma_{j[\sigma,\nu]}\equiv\left(\left\langle \sigma_{z}\right\rangle \tau_{x},-\tau_{y},\left\langle \nu_{z}\right\rangle \tau_{z}\right)$,
i.e., the four Weyl nodes are eigenstates of $\sigma_{z}$ and $\nu_{z}$
so that their $2\times2$ Weyl Hamiltonians can simply be written
by replacing $\sigma_{z}$ and $\nu_{z}$ in (\ref{eq:linearized H})
by their eigenvalues. Additionally, momenta have been rescaled to
make the nodes isotropic, for convenience, and the Dirac velocity has been set to unity. Note that momentum is now
measured from the node. $H_{0}$ is the most general Hamiltonian for
four degenerate Weyl nodes, because any other suitable Hamiltonian
is unitarily related to $H_{0}$ and can differ only in the orbital
content of the Dirac matrices.\emph{ }\textit{\emph{Thus, the results
obtained here can be straightforwardly extended to other similar systems
by appropriate unitary transformations}}\emph{.}

Assuming $\mu>0$ without loss of generality, the wavefunctions for
the Fermi level states $|\boldsymbol{k},j[\sigma_{z},\nu_{z}]\rangle\equiv|\boldsymbol{k},\left\langle \sigma_{z}\right\rangle ,\left\langle \nu_{z}\right\rangle \rangle$
are 
\begin{eqnarray}
|\boldsymbol{k},-,+\rangle & \equiv & \left(\begin{array}{c}
1\\
0
\end{array}\right)\otimes\left(-\cos\frac{\theta}{2},e^{i\phi}\sin\frac{\theta}{2}\right)^{T}\otimes\left(\begin{array}{c}
0\\
1
\end{array}\right)\nonumber \\
|\boldsymbol{k},+,-\rangle & \equiv & i\nu_{x}\sigma_{y}\mathbb{K}|-\boldsymbol{k},-+\rangle\nonumber \\
|\boldsymbol{k},-,-\rangle & \equiv & \nu_{x}\tau_{z}|-\boldsymbol{k},-+\rangle\nonumber \\
|\boldsymbol{k},+,+\rangle & \equiv & i\nu_{x}\sigma_{y}\mathbb{K}|-\boldsymbol{k},--\rangle\label{eq:spinors}
\end{eqnarray}
where the first, second and third factors in the direct product refer
to the valley ($\nu$), orbital ($\tau$) and spin ($\sigma$) parts
of the spinors, respectively, and $\theta$ and $\phi$ are the usual
spherical polar angles of $\boldsymbol{k}$. These spinors will be
used to project any interaction onto the FS.

\subsection{First order\label{sub:First-order}}

To first order in the interactions, the calculation is straightforward
and can be done analytically. We simply project each interaction onto
FS states related by $\mathcal{T}$ or $\mathcal{I}$ and obtain the
effective interaction matrix $\chi_{\mathcal{O},ij}^{X}(\boldsymbol{k},\boldsymbol{k}')=\langle\boldsymbol{k}',j|\otimes\langle\mathcal{O}(\boldsymbol{k}',j)|X|\mathcal{O}(\boldsymbol{k},i)\rangle\otimes|\boldsymbol{k},i\rangle$
where $i\in\{\pm1,\pm2\}$, $X$ is one of the two-body interaction
Hamiltonian operators listed in Sec \ref{sec:ABi-system} and $\mathcal{O}=\mathcal{I},\mathcal{T}$. Once we
have the total effective interaction for a particular type of Cooper
pairs, $\chi_{\mathcal{O}}(\boldsymbol{k},\boldsymbol{k}')=\sum_{X}\chi_{\mathcal{O}}^{X}(\boldsymbol{k},\boldsymbol{k}')$,
the pairing instabilities are determined by solving the eigenvalue
equations (\ref{eq:eigenvalue-equation}) and (\ref{eq:eigenvalue-equation-1}).
The results to first order are as follows.

\subsubsection{Hubbard interaction and pair-hopping:}

For simplicity, we first present results for $U_{S}=U_{P}=U$. Then, the Hubbard interaction
and pair-hopping have the same set of eigenstates for (\ref{eq:eigenvalue-equation})
and can be studied together. The effective interaction for non-zero
$U$ and $W$ but $V=J_{z}=0$ for type-T Cooper pairs is 
\begin{multline}
\chi_{\mathcal{T}}^{U}(\boldsymbol{k},\boldsymbol{k}')+\chi_{\mathcal{T}}^{W}(\boldsymbol{k},\boldsymbol{k}')=\frac{U+W}{8}\left(\begin{array}{cccc}
1 & 1 & 1 & 1\\
1 & 1 & 1 & 1\\
1 & 1 & 1 & 1\\
1 & 1 & 1 & 1
\end{array}\right)\\
+\frac{U-W}{8}\cos\theta\cos\theta'\left(\begin{array}{cccc}
1 & -1 & 1 & -1\\
-1 & 1 & -1 & 1\\
1 & -1 & 1 & -1\\
-1 & 1 & -1 & 1
\end{array}\right)
\end{multline}
This has two sets of eigenstates with non-trivial eigenvalues. They
are 
\begin{eqnarray}
\Delta_{S}(\boldsymbol{k}) & = & (1,1,1,1)^{T},~\lambda_{0}=\frac{U+W}{2}\\
\Delta_{P,n}(\boldsymbol{k}) & = & \cos^{2n-1}\theta(1,-1,1,-1)^{T},~\lambda_{n}=\frac{U-W}{4n+2}\nonumber 
\end{eqnarray}
where $n\in\mathbb{Z}>0$. All these states preserve $\mathcal{T}$
and $\mathcal{I}$. $\Delta_{S}$ is gapped while $\Delta_{P,n}$
has a line node at $\theta=\pi/2$. Clearly, a pairing instability
exists in one of the above channels when $|W|>U$. If $U_{S}\neq U_{P}$,
the gap functions develop additional variations over the FSs, but
the phase remains the same. Only in the extreme case where one of
$U_{S}$ and $U_{P}$ vanishes, the gap functions develop additional
nodes.

For type-I Cooper pairs, the effective interaction in these channels
vanishes. This is because the creation (or annihilation) operators
in these interactions are Kramer's conjugates and thus have opposite
spins, but type-I Cooper pairs involve equal spin pairing. Thus, projecting
the Hubbard or pair-hopping interaction onto states related by $\mathcal{I}$
identically gives zero.

\subsubsection{Inter-orbital repulsion and Ising exchange:}

These two interactions have the same eigenstates for (\ref{eq:eigenvalue-equation}) or (\ref{eq:eigenvalue-equation-1})
and can be conveniently studied together. For type-T Cooper pairs,
the effective interaction for $U=W=0$ and non-zero $V$ and $J_{z}$
is 
\begin{multline}
\chi_{\mathcal{T}}^{V}(\boldsymbol{k},\boldsymbol{k}')+\chi_{\mathcal{T}}^{J_{z}}(\boldsymbol{k},\boldsymbol{k}')=\frac{V-J_{z}}{8}\sin\theta\sin\theta'\times\\
\cos(\phi-\phi')\left(\begin{array}{cccc}
1 & -1 & -1 & 1\\
-1 & 1 & 1 & -1\\
-1 & 1 & 1 & -1\\
1 & -1 & -1 & 1
\end{array}\right)
\end{multline}
Clearly, only the difference $V-J_{z}$ is important for type-T Cooper
pairs. The eigenstates with non-trivial eigenvalues are 
\begin{equation}
\Delta_{\pm}^{\mathcal{T}}(\boldsymbol{k})=\sin\theta e^{\pm i\phi}(1,-1,-1,1)^{T}\,,\,\lambda_{\pm}^{\mathcal{T}}=\frac{V-J_{z}}{6}
\end{equation}
These states have point nodes at $\theta=0,\pi$ and are odd under
$\mathcal{I}$ and break $\mathcal{T}$ . A pairing instability exists
in this channel if $J_{z}>V$.

For type-I Cooper pairs, the effective interaction is 
\begin{multline}
\chi_{\mathcal{I}}^{V}+\chi_{\mathcal{I}}^{J_{z}}=\frac{V+J_{z}}{4}\sin\theta\sin\theta'\times\\
\left(\begin{array}{cccc}
e^{i(\phi-\phi')} & 0 & 0 & e^{i(\phi-\phi')}\\
0 & e^{-i(\phi-\phi')} & e^{-i(\phi-\phi')} & 0\\
0 & e^{-i(\phi-\phi')} & e^{-i(\phi-\phi')} & 0\\
e^{i(\phi-\phi')} & 0 & 0 & e^{i(\phi-\phi')}
\end{array}\right)
\end{multline}
and has the spectrum 
\begin{eqnarray}
\Delta_{+}^{I}(\boldsymbol{k})=\sin\theta e^{i\phi'}\left(\begin{array}{cccc}
1 & 0 & 0 & 1\end{array}\right)^{T} & , & \lambda_{+}^{I}=\frac{V+J_{z}}{3}\\
\Delta_{-}^{I}(\boldsymbol{k})=\sin\theta e^{-i\phi'}\left(\begin{array}{cccc}
0 & 1 & 1 & 0\end{array}\right)^{T} & , & \lambda_{-}^{I}=\frac{V+J_{z}}{3}
\end{eqnarray}
These states are odd under $\mathcal{I}$, break $\mathcal{T}$-symmetry
and have nodes at $\theta=0,\pi$. They appear when $J_{z}<-V$.

Thus, there exist various gapped as well as nodal phases. However,
the TSC, given by $\Delta\sim(1,1,-1,-1)^{T}$, is an eigenstate of
each interaction with eigenvalue zero. In other words, there is no
energy gain in forming the TSC, as we had anticipated in Sec \ref{sec:microscopic-origins}.

\subsection{Second order}

\begin{figure}[h]
\vspace{12pt}
\subfloat{%
\begin{tabular}{|c|c|}
\hline 
\includegraphics[width=0.45\columnwidth]{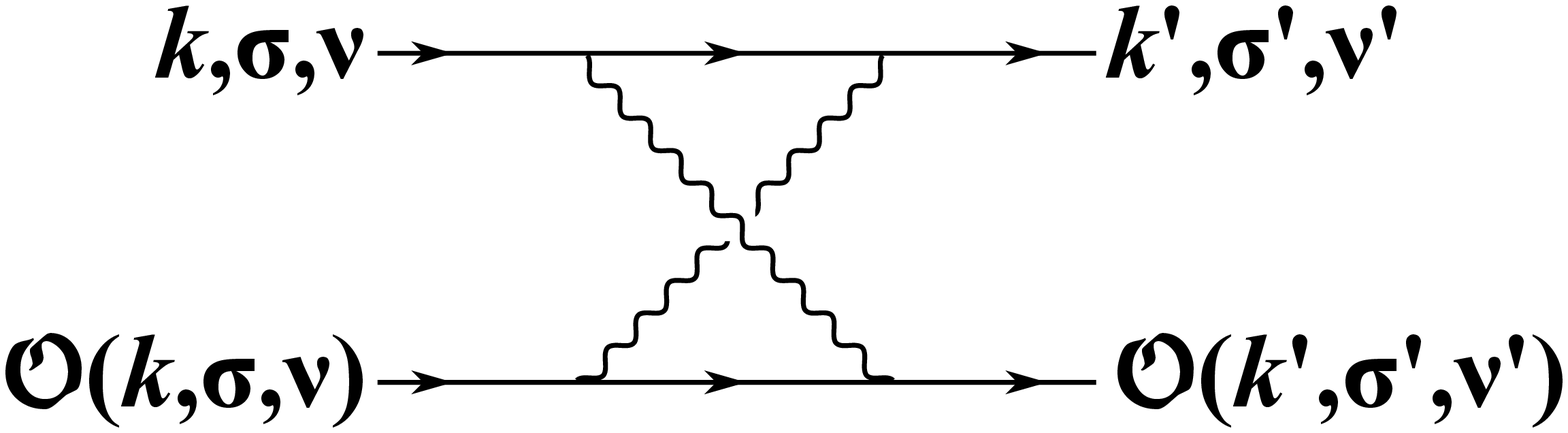}  & \includegraphics[width=0.45\columnwidth]{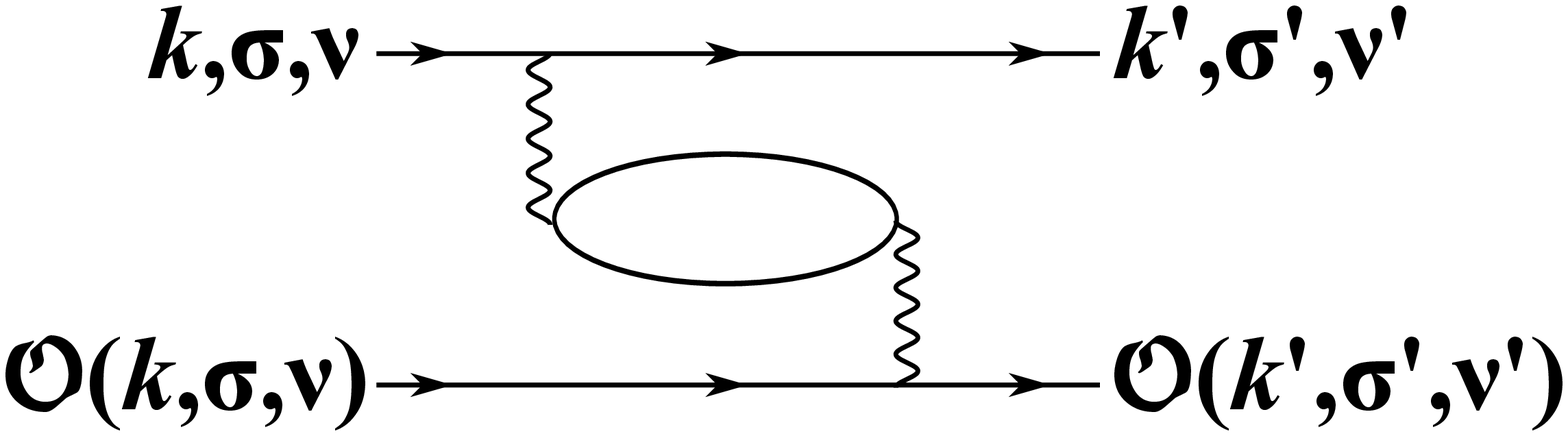}\tabularnewline
\hline 
\includegraphics[width=0.45\columnwidth]{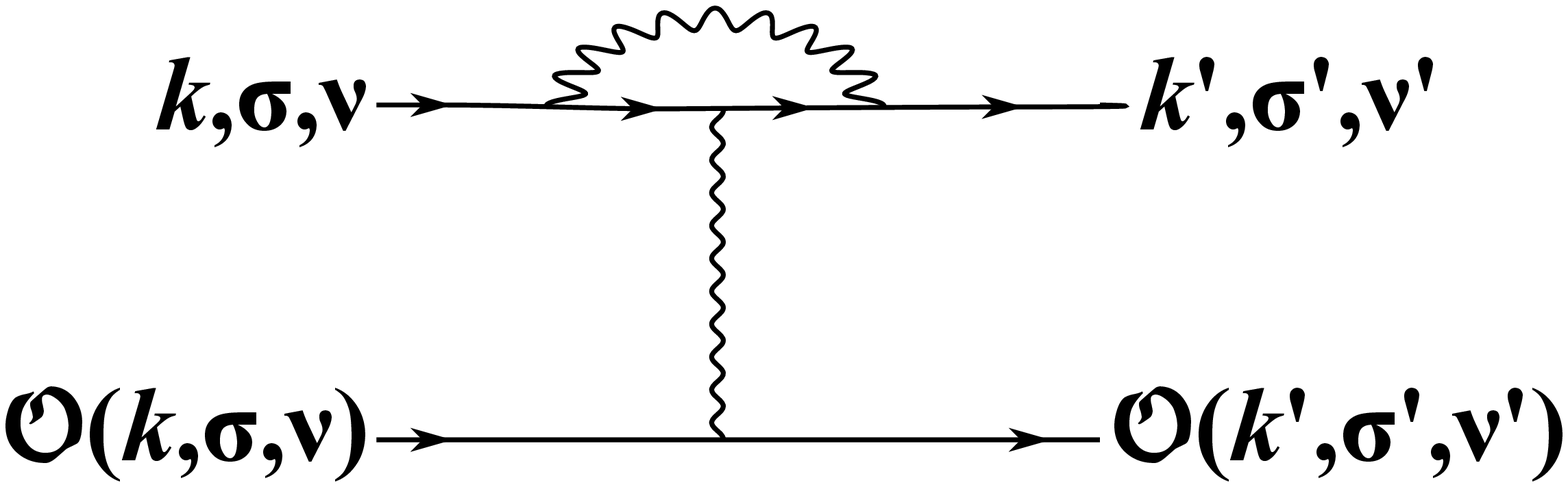}  & \includegraphics[width=0.45\columnwidth]{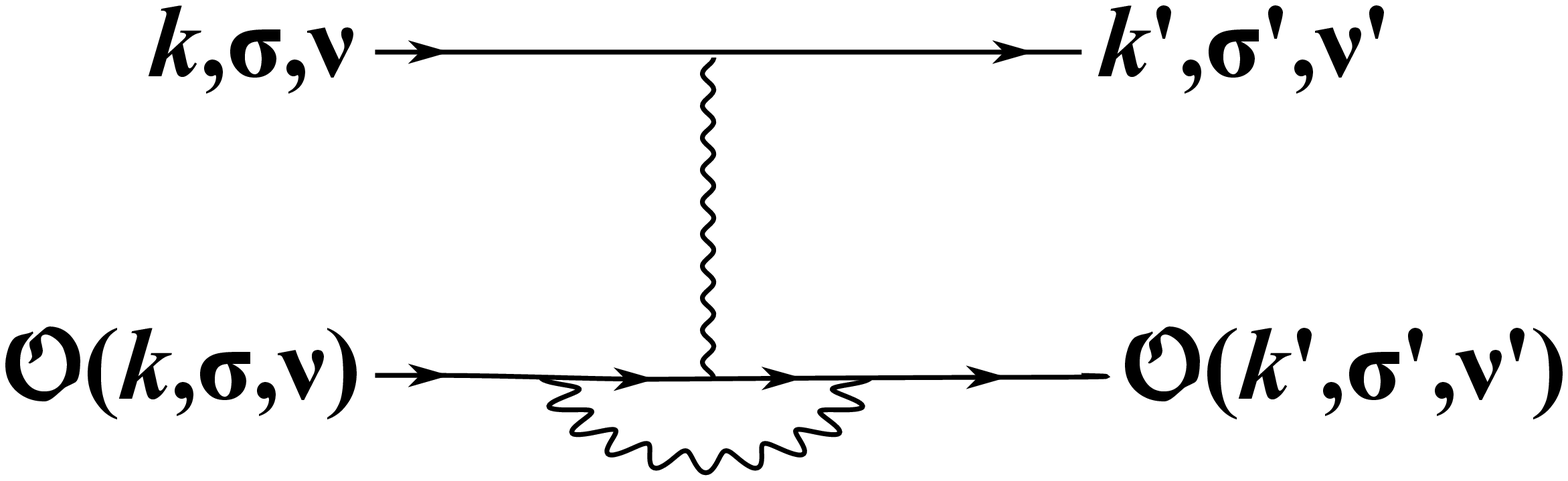}\tabularnewline
\hline 
\end{tabular}}

\caption{Diagrams that contribute to the effective interaction to second order
in the bare interactions. The wavy lines represent any one of the
five interactions $U_{s}$, $U_{p}$, $V$, $J_{z}$ and $W$. \label{fig:second-order-diagrams}}
\end{figure}

We now consider two-particle scattering processes in the Cooper channel
to second order in the bare interactions. The diagrams to this order
are depicted in Fig \ref{fig:second-order-diagrams}. As usual, the
solid lines represent fermions and each wavy line denotes any one
of the five interactions $U_{s}$, $U_{p}$, $V$, $J_{z}$ and $W$.

These processes introduce momentum dependence into the interaction.
The first two of these processes have been studied in the past for
electrons interacting purely via Hubbard repulsion. While the former
has been shown to induce spin singlet $d$-wave superconductivity
near a spin density wave instability\cite{ScalapinoDWave}, the latter
mediates ferromagnetic fluctuations and induces an instability towards
a $p$-wave superconductor with equal spin pairing\cite{RaghuWeakRepulsiveSC}.
The bottom row of diagrams does not exist for the Hubbard interactions
$U_{s,p}$ or pair-hopping $W$, but does, for $V$ and $J_{z}$.
It can be checked that these four diagrams are related to one another
by an exchange of a pair of fermion operators for one or both of the
bare interactions. Integrating over the internal lines gives\begin{widetext}
\begin{equation}
\lim_{\beta\to\infty}\frac{1}{\beta}\sum_{i\omega_{n}}\int\frac{\mathrm{d}\boldsymbol{k}^{\prime\prime}}{(2\pi)^{3}}G_{ab}^{j_{1}}(i\omega_{n},\boldsymbol{k}^{\prime\prime})G_{cd}^{j_{2}}(i\omega_{n},\boldsymbol{k}^{\prime\prime}-\boldsymbol{Q})=\begin{cases}
-\frac{\Lambda^{2}}{16\pi^{2}}\left[\delta_{ab}\delta_{cd}-\frac{1}{3}\boldsymbol{\Gamma}_{ab}^{j_{1}}\cdot\boldsymbol{\Gamma}_{cd}^{j_{2}}\right] & Q=0\\
-\frac{\Lambda^{2}}{16\pi^{2}}\left[\delta_{ab}\delta_{cd}+\boldsymbol{\Gamma}_{ab}^{j_{1}}\cdot\boldsymbol{\Gamma}_{cd}^{j_{2}}-2(\boldsymbol{\Gamma}_{ab}^{j_{1}}\cdot\hat{\boldsymbol{Q}})(\boldsymbol{\Gamma}_{cd}^{j_{2}}\cdot\hat{\boldsymbol{Q}})\right] & Q\neq0
\end{cases}\label{eq:GreenCrossSum}
\end{equation}
\end{widetext}to lowest order in $\mu/\Lambda$. Here, $G_{ab}^{j}(i\omega,\boldsymbol{k})=\left[i\omega-\boldsymbol{\Gamma}_{j}\cdot\boldsymbol{k}+\mu\right]_{ab}^{-1}$
is the Matsubara Green's function of an internal fermion line, $(i\omega_{n},\boldsymbol{k},j)$
are the internal frequency, momentum relative to the Weyl node and
Weyl node index, respectively, $\Lambda$ is a momentum cut-off which
physically corresponds to the scale where non-linearities of the dispersion
become important and $\boldsymbol{Q}$ is the difference in momenta
of the internal lines and takes one of the four values $\pm\boldsymbol{k}\pm\boldsymbol{k}'$
depending on the diagram.

The rank-4 tensors in (\ref{eq:GreenCrossSum}) are contracted with
the two rank-4 interaction tensors denoted by the wavy lines in Fig \ref{fig:second-order-diagrams}, the
resulting expression is anti-symmetrized with respect to the external
lines, and finally projected onto FS states using (\ref{eq:spinors}).
Although the integrals can be done analytically, the bookkeeping of
the diagrams in the presence of all the interactions is best done
numerically. Finally, the first order interactions are added to the
results to obtain $\chi_{\mathcal{O}}(\boldsymbol{k},\boldsymbol{k}')$
for each type of Cooper pairs, and the eigenvalue equations (\ref{eq:eigenvalue-equation})
and (\ref{eq:eigenvalue-equation-1}) are solved to determine the
superconducting phases. The results are discussed in next subsection.

\subsection{Summary of the results}

\begin{figure}
\begin{centering}
\includegraphics[width=0.9\columnwidth]{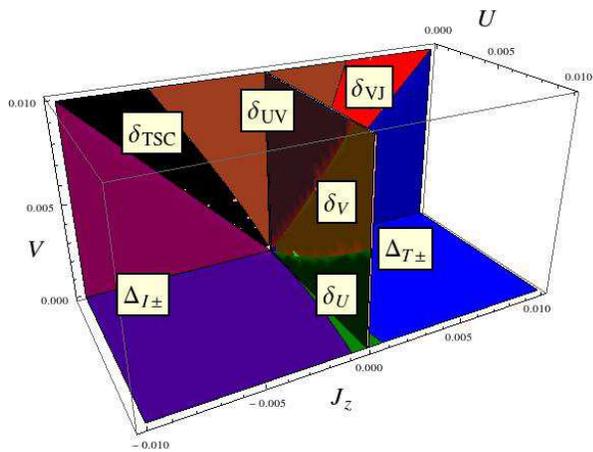} 
\par\end{centering}

\caption{Phase diagram upto second order in the interactions at $W=0$ in the
presence of $\mathcal{I}$-symmetry. The gray phase at negative $J_{z}$
and intermediate $U$ or $V$ is the $\mathcal{T}$-symmetric topological
superconductor. The phases were determined numerically, using $\mu=0.1$
and $\Lambda=1.0$. Details of all the phases are in Table \ref{tab:Phases}.
\label{fig:Phase-diagram-second-order-I}}
\end{figure}

\begin{figure}
\begin{centering}
\includegraphics[width=0.9\columnwidth]{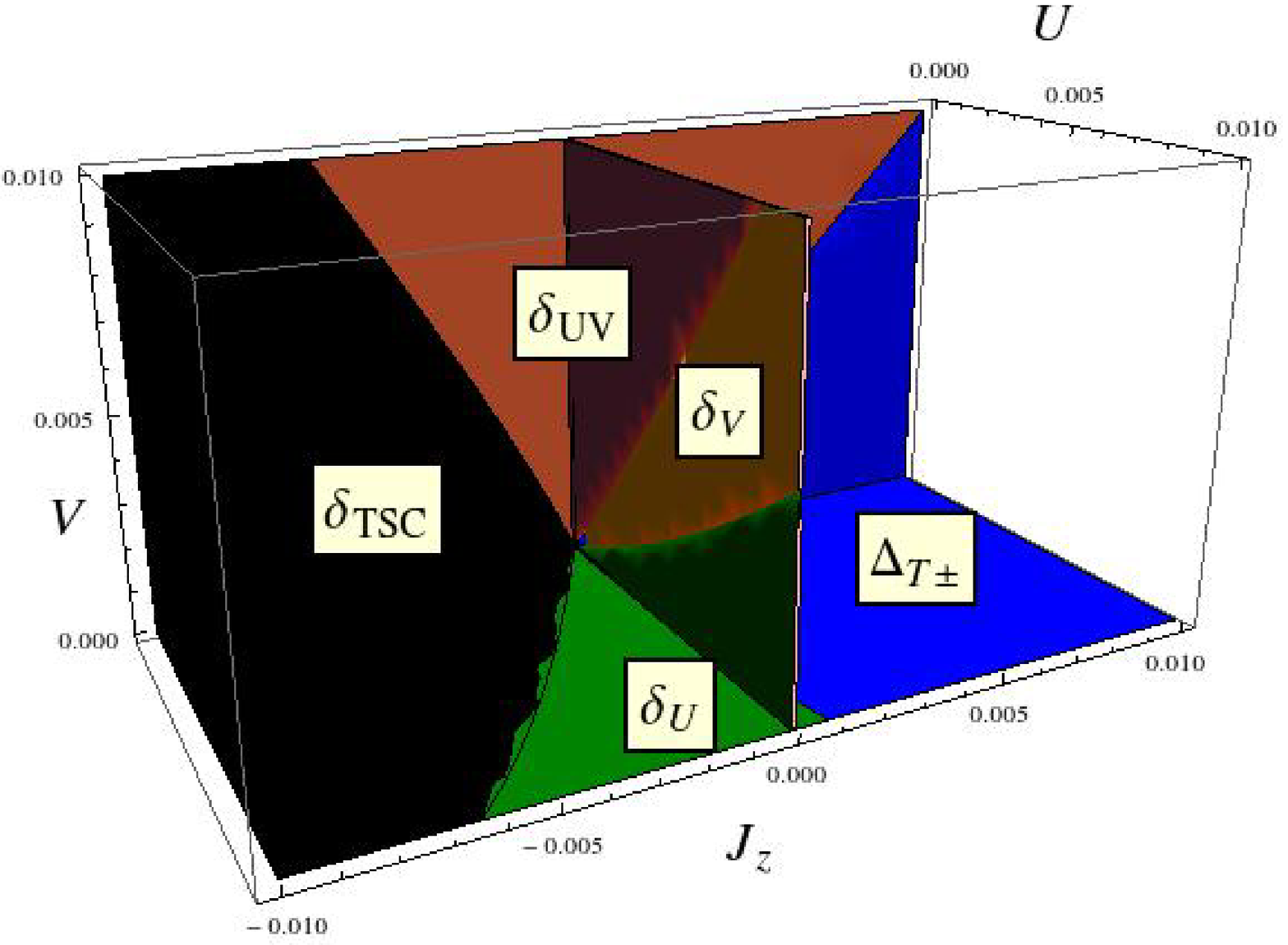} 
\par\end{centering}

\caption{Phase diagram upto second order in the interactions at $W=0$ in the
absence of $\mathcal{I}$-symmetry. The black phase at negative $J_{z}$
is the $\mathcal{T}$-symmetric TSC, which occupies a larger part
of the phase diagram compared to Fig \ref{fig:Phase-diagram-second-order-I}
when $\mathcal{I}$-symmetry is present. The phases were determined
numerically, using $\mu=0.1$ and $\Lambda=1.0$. The $J_{z}=0$ cross-section
is identical to the one in Fig \ref{fig:Phase-diagram-second-order-I}.
Details of all the phases are in Table \ref{tab:Phases}.\label{fig:Phase-diagram-second-order-T}}
\end{figure}

The fluctuation-exchange approach discussed so far can be applied
to systems with generic interaction parameters. To understand the
result, we make the choice $U_{s}=U_{p}\equiv U$ and $W=0$ and present
the phase diagram in the space of $J_{z},V,U$ in Figs. \ref{fig:Phase-diagram-second-order-I}
and \ref{fig:Phase-diagram-second-order-T}. $W=0$ is chosen on grounds
of realizability since it is the least common of all interactions
considered here. A small non-zero $W$ clearly does not affect the
type-I pairing states to first order because they involve equal spin
pairing whereas $W$ involves hopping of Kramers conjugate states
which have opposite spin. For type-T pairing states, a straightforward
evaluation of the matrix elements of $\chi_{\mathcal{T}}^{W}$ in
the basis of the eigenvectors describing each state shows that $\delta_{U}$
and $\delta_{V}$ are destabilised by $W$ while the other states
are unaffected to first order in $W$.

\begin{table}
\begin{centering}
\begin{tabular}{|c|c|c|c|c|c|}
\hline 
Phase  & Cooper pair  & $\mathcal{T}$  & Nodal structure  & $\phi$-dependence\tabularnewline
\hline 
\hline 
$\Delta_{T\pm}$  & T-type  & $\times$  & 2 point nodes  & $e^{\pm i\phi}$ \tabularnewline
\hline 
$\Delta_{I\pm}$  & I-type  & $\times$  & 2 point nodes  & $e^{\pm i\phi}$\tabularnewline
\hline 
\textbf{$\delta_{TSC}$}  & \textbf{T-type}  & \textbf{$\boldsymbol{\checkmark}$}  & \textbf{gapped}  & \textbf{none}\tabularnewline
\hline 
$\delta_{U}$  & T-type  & $\checkmark$  & 2 line nodes  & none\tabularnewline
\hline 
$\delta_{V}$  & T-type  & $\times$  & 2 point, 1 line  & $e^{\pm i\phi}$\tabularnewline
\hline 
$\delta_{UV}$  & T-type  & $\times$  & 2 point nodes  & $e^{\pm2i\phi}$\tabularnewline
\hline 
$\delta_{VJ}$  & I-type  & $\times$  & 2 point, 2 line  & $e^{\pm i\phi}$ \tabularnewline
\hline 
\end{tabular}
\par\end{centering}

\caption{Details of the phases in Figs. \ref{fig:Phase-diagram-second-order-I}
and \ref{fig:Phase-diagram-second-order-T}. $\Delta$-phases ($\delta$-phases)
appear to first (second) order in the interactions. $\phi$ is the
azimuthal angle in momentum space. The phases with no $phi$-dependence are non-degenerate while the rest are doubly degenerate, with one state corresponding to each sign in the `$phi$-dependence' column.  $\delta_{TSC}$ is the topological
superconductor.\label{tab:Phases}}
\end{table}

The black phase labeled $\delta_{TSC}$ is the TSC. When $\mathcal{I}$-symmetry
is preserved, $\delta_{TSC}$ appears in a narrow range of ferromagnetic
$J_{z}$ and repulsive $V$ , and is dominated by a nodal phase if
either of these interactions is enhanced. Importantly, the nodal phase
at large ferromagnetic $J_{z}$ is an equal spin pairing state, consisting
of type-I Cooper pairing, and is thus precluded if $\mathcal{I}$-symmetry
is broken. In the absence of $\mathcal{I}$-symmetry, therefore, $\delta_{TSC}$
appears over a much larger region of the phase diagram. In particular,
it forms if interactions are purely ferromagnetic, and survives moderate
$V$ and $U$.

We make a quick note about a subtlety of the type-I pairing states.
Since the FSs have Chern numbers, the wavefunctions on them cannot
be defined globally. For any phase choice of the wavefunction, each
FS has a point around which the phase of the wavefunction winds, but
the magnitude is non-vanishing. Consequently, the gap functions projected
onto the Fermi level are not smooth either even though the Cooper
pair wavefunctions written in terms of the underlying orbitals are.
This is not an issue for the type-T Cooper pairs, because the phases
of wavefunctions of Kramer's conjugates mutually cancel.

\section{Conclusion and discussions}

In summary, we have proposed a general procedure for the search of
$\mathcal{T}$-invariant topological superconductivity in doped WSMs. The nontrivial Berry
curvature at the FSs of WSMs allows the TSC to be realized for
a pairing function with no special momentum dependence like in He$^{3}$B.
Using the fluctuation-exchange approach, we discovered the general
requirements on the effective interaction for the realization of a
TSC phase using a minimal model of $\mathcal{T}$-invariant WSM. As
an explicit example, we study a prototype model describing Dirac semimetal
${\rm A_{3}Bi}$ and its $\mathcal{I}$-breaking deformation. We studied
explicitly the possible interaction terms up to second order of the
bare interaction, and obtained the phase diagram of possible superconductor
phases. We showed that the TSC exists when the ferromagnetic exchange
coupling is greater than or comparable to the repulsive density-density
interactions, and is stabilised greatly by the violation of $\mathcal{I}$-symmetry. Thus, our results suggest searching for the TSC phase in doped
${\rm A_{3}Bi}$ family of materials. Interestingly, another Dirac
semimetal, ${\rm Cd_{3}As_{2}}$, was theoretically predicted\cite{wang2013}
and experimentally realized\cite{he2014,jeon2014,liang2014,neupane2014}
recently. Different from ${\rm A_{3}Bi}$, ${\rm Cd_{3}As_{2}}$ is
proposed to be $\mathcal{I}$-breaking, although the $\mathcal{I}$-breaking
does not separate the Weyl points, but instead induces a velocity
anisotropy\cite{wang2013}. Nonetheless, it would be extremely interesting
to investigate what kinds of superconductivity are possible in this
material; in particular, is it a better candidate for possible TSCs
than ${\rm A_{3}Bi}$, since it already breaks inversion symmetry?

Although we used the prototype model as an example, our proposal of
realizing TSC in doped WSM is generic and can apply to other doped
WSM systems with suitable electronic interactions. The general idea
is that the topological nontrivial FSs in doped WSM already provide
a suitable normal state for TSC, and the inversion symmetry breaking
helps to suppress other competing superconducting orders. For a given
doped WSM system, the fluctuation-exchange method can be used to determine
whether the electron interaction in this system prefers TSC. Now that
the list of WSMs and Dirac semimetals is growing rapidly, our proposal
provides a guiding principle and a general method for the search of
a 3D TSC in this large family of materials. An interesting future
direction is to generalize this approach to systems with degenerate
points on the FSs, \textit{i.e.}, touching points of multiple FSs.
In such cases, the pairing order parameter can generically take a
matrix form at the degenerate points, allowing richer possibilities
for superconductivity.

\begin{acknowledgments} PH would like to thank Srinivas Raghu for helpful discussions and David and Lucile
Packard Foundation as well as the Department of Energy Office of Basic Energy Sciences,
contract DE-AC02-76SF00515 for financial support. XD and ZF acknowledge support from the National
Science Foundation of China and the 973 Program of China (No. 2011CBA00108
and No. 2013CB921700). XLQ is supported by the National Science Foundation
through the grant No. DMR-1151786.
\end{acknowledgments}

\bibliographystyle{apsrev4-1}
\bibliography{references}

\end{document}